\documentclass[aps,prb,twocolumn,showpacs,amsmath,amssymb,floatfix,superscriptaddress]{revtex4}
\usepackage{graphicx}
\usepackage{subfigure}
\newcommand {\be}{\begin{equation}}
\newcommand {\ee}{\end{equation}}

\begin{document}

\title{Heat wave propagation in a nonlinear chain}
\date{\today}

\author{Francesco Piazza}
\email{Francesco.Piazza@epfl.ch}
\affiliation{Laboratoire de Biophysique Statistique, Institut de th\'eorie 
des ph\'enom\`enes physiques,
Ecole Polytechnique F\'ed\'erale de Lausanne (EPFL)
BSP-722, CH-1015
Lausanne, Switzerland}

\author{Stefano Lepri}
\email{stefano.lepri@isc.cnr.it}
\affiliation{Istituto dei Sistemi Complessi, Consiglio Nazionale
delle Ricerche, via Madonna del Piano 10, I-50019 Sesto Fiorentino, Italy}

\begin{abstract}
We investigate the propagation of temperature perturbations
in an array of coupled nonlinear oscillators at finite 
temperature. We evaluate the response function at equilibrium 
and show how the memory effects affect the diffusion
properties. A comparison with nonequilibrium simulations
reveals that the telegraph equation provides a reliable 
interpretative paradigm for describing 
quantitatively the propagation of a heat  pulse at the 
macroscopic level. The results could be of help in understanding
and modeling energy transport in individual nanotubes. 
\end{abstract}

%
%
% 05.45.-a 	Nonlinear dynamics and chaos 
% 05.60.-k 	Transport processes
%
%      05.60.Cd 	Classical transport
%      05.60.Gg 	Quantum transport
%
% 44.10.+i 	Heat conduction (see also 66.25.+g and 66.70.?f in nonelectronic transport properties of condensed matter)
% 66.70.-f 	Nonelectronic thermal conduction and heat-pulse propagation in solids; thermal waves (for electronic thermal 
%           conduction in metals and alloys, see 72.15.Cz and 72.15.Eb)
% 66.10.C- 	Diffusion and thermal diffusion (for osmosis in biological systems, see 82.39.Wj in physical chemistry; 
%           for cellular transport, see 87.16.dp and 87.16.Uv in biological physics)
% 66.10.cd 	Thermal diffusion and diffusive energy transport
% 
%

\maketitle

\section{Introduction}

The propagation of temperature fluctuations under nonequilibrium 
conditions may display significant deviations from a simple 
diffusive behavior. As it is know ~\cite{preziosi}, heat can propagate 
in a wavelike form on time scales comparable with some 
typical relaxation time. This is accounted for by suitable extensions
of the standard Fourier's law to include, for 
example, a finite response time for the heat current.   
On the other hand, it is of primary interest to derive such 
a phenomenological laws from microscopic dynamics. In this 
respect, much work has been performed recently, mostly on 
simplified models, with the goal of understanding nonequilibrium
heat--conducting states from first principles ~\cite{LLP03}.
Much emphasis has been put on violations of Fourier's law in the 
stationary case ~\cite{LLP03} as well as in describing anomalous 
heat diffusion ~\cite{denisov}. Transient heat propagation 
has received much less attention ~\cite{zhao,Delfini07a}.
As a matter of fact, the spreading of the energy perturbation field,
yield complementary information on how heat propagates 
through the system \cite{Helfand60} and provides useful 
insight on the nature of heat carrying excitations.

Besides those fundamental issues,
there is a growing interest in understanding how heat is 
transported at the nanoscale ~\cite{cahill}. In this context,
carbon nanotube materials are of special relevance due to their 
exceptionally high thermal conductivity related to their 
quasi one-dimensional vibrational structure ~\cite{hone,Yu05}. 
Recently, Osman and Srivastava ~\cite{osman} investigated the 
propagation of intense heat pulses in single--walled carbon nanotubes
by means of molecular dynamics. They observed that, together with 
pulses traveling with the sound velocity, a secondary and slower peak
can also propagate as a "second sound" type of 
wave ~\cite{Chester}.
Shiomi and Maruyama~\cite{maruyama}, argued that this should 
be related to relatively fast optical phonons, that due to the
quasi one--dimensional structure, have very long relaxation
times and thus contribute to wavelike heat propagation.

It is thus relevant to study simplified models that can help
to understand better the conduction properties. 
In this respect, an example is the length dependence
of conductivity in carbon and boron-nitride nanotubes which 
has been very recently observed experimentally \cite{Chang08}.
Indeed, experimental data are compatible with scaling laws
theoretically predicted for simple nonlinear models.
In this spirit, in the present paper we analyze the 
heat pulse propagation in a classical lattice model, a chain
of classical, harmonically coupled, oscillators with a 
quartic pinning (on-site) potential. In particular, we compare 
the linear-response predictions with nonequilibrium molecular
dynamics.

The paper is organized as follows. In Sec. \ref{sec:model} we introduce the 
microscopic one-dimensional model. After some general considerations about
linear response we present the response functions as computed from molecular
dynamics simulations (Sec. \ref{sec:resp}). Based on the  numerical data, we
introduce an approximated (single-pole) form for the response that leads to the
telegraph equation for the temperature--field evolution. This equation, along 
the phenomenological values of its parameters, is precious to compare  with
nonequilibrium simulations. The transient evolution of a temperature pulse
is  investigated in Sec. \ref{sec:pulse}. Qualitative and quantitative 
deviations from the behavior expected for the simple telegraph equation
are highlighted. 

\section{The nonlinear chain}
\label{sec:model}

The model consists of an anharmonic chain of $N$ particles (each with 
unit mass) whose displacements are denoted by $u_i$:  
\begin{equation}
\ddot u_i = - u_i -u_i^3 + C (u_{i+1} - 2u_i + u_{i-1}) \quad.
\label{eqmot}
\end{equation}
In the linear approximation, where the cubic force term 
is dropped, the eigenfrequencies $\Omega$ of the associated 
normal modes are expressed as a function of the wavenumber $q$ 
by
\begin{equation}
\Omega^2(q) \;=\; 1 + 2C(1 -\cos q)\quad.
\end{equation}
The units have been fixed in such a way to have a unitary gap in the 
spectrum. With this choice the only free parameters of the model are the 
energy per particle $\epsilon=E/N$ and the coupling constant $C$.

Several numerical studies (see e.g. Ref.~\onlinecite{Hu98} and 
\onlinecite{aoki} as well as Ref. ~\onlinecite{LLP03} and the bibliography
within) clarified, that for models like (\ref{eqmot}) the thermal 
conductivity is finite, and that a diffusive heat propagation 
is expected at long times. However, on time scales which are 
of the order of the energy current relaxation time some wave-like 
transient behavior could be expected.  However, a rigorous derivation
of a macroscopic transport law from the microscopic equations of motion, 
especially for small 
systems in a low-dimensional environment, remains a formidable challenge.

To conclude this Section we mention that the we limit ourselves
here to the case of a single-well on-site potential.  
The double-well case has been studied in Ref.~\onlinecite{flach}.
A detailed analysis of second sound propagation in the 
three-dimensional lattice is given in Ref.~\onlinecite{stoll}.

\section{Linear response}
\label{sec:resp}

At a coarse-grained level, 
the most general linear constitutive relation between the energy current
$J(x,t)$ and the local temperature gradient can be written as 
an extension of the usual Fourier law (we limit ourselves 
for simplicity to a one-dimensional case),
\begin{equation}
J(x,t) \;=\; - \int_{-\infty}^{\infty} dx' \int_{-\infty}^t dt' K(x-x',t-t') 
\frac{\partial T}{\partial x}(x',t')  \quad.  
\label{fourier}
\end{equation}
Substituting into the continuity equation, and assuming a decay 
of the kernel $K$ at infinity, one obtains
\begin{equation}
\frac{\partial T}{\partial t} 
\;=\;  \int_{-\infty}^{\infty} dx' \int_{-\infty}^t dt' K(x-x',t-t') 
\frac{\partial^2 T}{\partial x^2}(x',t')   
\label{diffgen}
\end{equation}
(up to the heat capacity). The generalized heat diffusion equation 
is formally solved by means of the Fourier-Laplace transform, with
the convention 
\begin{equation}
f(q,z) \equiv \frac{1}{\sqrt{2 \pi}}   \int_{-\infty}^\infty dx \int_0^\infty dt \, f(x,t) e^{-i(qx-zt)} \quad, 
\end{equation}
yielding
\begin{equation}
T(q,z) = \frac{T_0(q)}{-iz + q^2 K(q,z)}\quad,
\label{Tzq}
\end{equation}
with $T_0$ being the initial condition. The standard diffusive pole
is recovered for a constant $K(q,z)=D$, i.e. for an instantaneous 
response, with $D$ being the thermal diffusivity. The simplest 
improvement would be to include a single 
relaxation time $\tau$, and no spatial memory effects, namely an 
exponentially decaying kernel in time of the Cattaneo-Vernotte 
type ~\cite{preziosi}. In the above notation this means choosing 
a single-pole response of the form
\begin{equation}
K(q,z) \;=\; \frac{v^2}{-iz + 1/\tau} \quad.
\label{catt}
\end{equation}
The quantity $v$ has the physical dimensions of a velocity, 
is defined as ~\cite{preziosi}
\begin{equation}
v = \sqrt{\frac{D}{\tau}}\quad.
\label{v} 
\end{equation} 

Substituting approximation (\ref{catt}) into Eq.~(\ref{Tzq})
one can readily recognize that the resulting $T(q,z)$ is nothing but 
the Laplace transform of the equation
\begin{equation}
\frac{\partial^2 T}{\partial t^2} + \frac{1}{\tau}
\frac{\partial T}{\partial t} = 
v^2\frac{\partial^2 T}{\partial x^2} 
\quad .
\label{teleg}
\end{equation}
This is known as the telegraph equation ~\cite{preziosi}.  The equation
contains an additional term with  respect to the standard heat equation that
affects the solution for times of order $\tau$. Indeed, the
second-order derivative with respect to time leads to a finite velocity of
perturbations. For instance, for a pulse initially localized at the origin, 
$T=0$ at distances $|x|>vt$.  For $t \gg \tau$, ordinary diffusive
behavior is recovered~\cite{morse}. It should be also remembered, that 
Eq.~(\ref{teleg}) arises as a continuum limit of the persistent 
random walk (see Ref.~\onlinecite{weiss} and references within).

%%%%%%%%%%%%%%%%%%%%%%%%%%%%%%%%%%%%%%%%%%%%%%%%%%%%%%%%%%%%%%%%%%%%%%%%%%%%%%%%%%%%%%%%%%%
\begin{figure}
\begin{center}
\includegraphics[width=\columnwidth,clip]{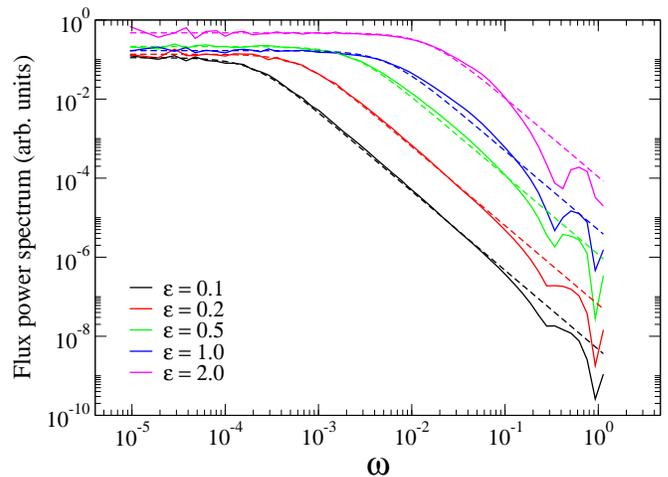}
\caption{Power spectrum of energy current for a chain 
of $N=4096$ $C=1$ and different energies. 
Dashed lines are best fit with a Lorentzian line-shape 
$a/(1+\tau^2\omega^2)$.
}
\label{spj}
\end{center}
\end{figure}
%%%%%%%%%%%%%%%%%%%%%%%%%%%%%%%%%%%%%%%%%%%%%%%%%%%%%%%%%%%%%%%%%%%%%%%%%%%%%%%%%%%%%%%%%%%

\begin{figure}
\begin{center}
\includegraphics[width=\columnwidth,clip]{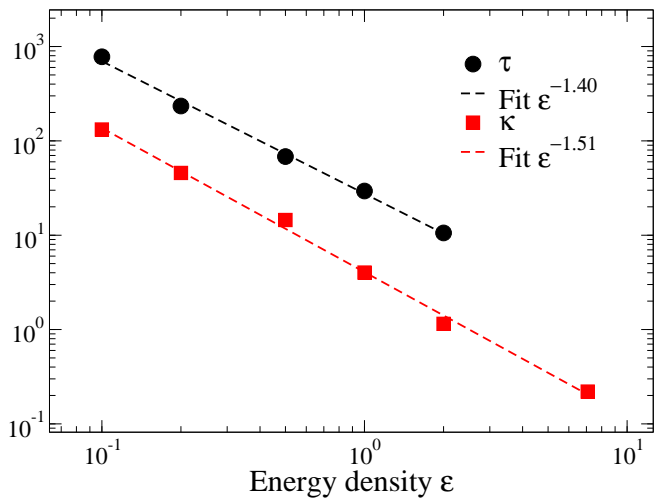}
\caption{The relaxation time $\tau$ as a function of energy
for $N=2048$, $C=1$. The dashed line is a power-law fit.
}
\label{times}
\end{center}
\end{figure}
According to the fluctuation--dissipation theorem ~\cite{lubensky}
the imaginary part of the response function $K(q,\omega)$ is 
proportional to $\omega$ times the equilibrium power spectrum of the
observable that couples to the external field. In the case of 
a temperature gradient, the observable to consider is 
the energy current, whose microscopic
expression for model (\ref{eqmot}) reads ~\cite{LLP03}
\begin{equation}
j_n = - \frac{C}{2}  (\dot u_{n+1} + \dot u_n)(u_{n+1} - u_n) \quad . 
\label{flx}
\end{equation}

In order to test for the validity of approximation (\ref{catt})
we have first of all evaluated the power spectra of the $q=0$ component
of the flux, $j(q=0,t)=\sum_n j_n$ from molecular dynamics. 
To this aim, we have performed microcanonical simulations by integrating
Eqs.~(\ref{eqmot}) (with periodic boundary conditions, $u_n=u_{n+N}$)  by means
of a fourth--order symplectic algorithm ~\cite{MA92}. Initial  conditions 
were chosen with the particles at equilibrium. Their velocities were
drawn at random  from a Gaussian distribution and rescaled by suitable factors
to assign the total energy per particle $\epsilon$ to the prescribed value and 
to set the total initial momentum equal to zero. A suitable transient is 
elapsed before data acquisition for statistical averaging. Conservation of energy 
and momentum was monitored during each run. The chosen time--steps 
(0.05--0.1) ensures energy conservation better then a few parts per million.    
The reliability of the spectra has been checked against different
choices of the run duration and sampling times. 

Fig.~\ref{spj} reports $\langle | j(q=0,\omega )|^2 \rangle$, where  the average
is performed over about 100  trajectories. The data  show that, in agreement
with Eq.~(\ref{catt}),  the spectra are fitted very well by a single Lorentzian
over a wide range of energies and frequencies. The  relaxation time decreases as
a power of the energy density, $\tau \sim \epsilon^{-\beta}$ with $\beta \approx
1.4$ (Fig.~\ref{times}). To compute the velocity $v$ we  measured
independently the thermal conductivity $\kappa$ by the standard Green-Kubo 
formula i.e. by integrating the flux autocorrelation in the time domain 
\cite{LLP03}. It turns out that the diffusivity $D$ is approximately  equal to
the thermal conductivity since the heat capacity is very close to one in our
units (it varies only by a few percent in the considered  energy range). As
shown again in Fig.~\ref{times}, $\kappa$ decreases as a power of the energy
density. This is in agreement with previous work on related
models ~\cite{aoki,Spohn06}. The corresponding exponent is very close to 
$\beta$ meaning the that $v$ as defined by Eq.~(\ref{v}) is roughly constant.

%%%%%%%%%%%%%%%%%%%%%%%%%%%%%%%%%%%%%%%%%%%%%%%%%%%%%%%%%%%%%%%%%%%%%%%%%%%%%%%%%%%%%%%%%%%
\begin{figure}
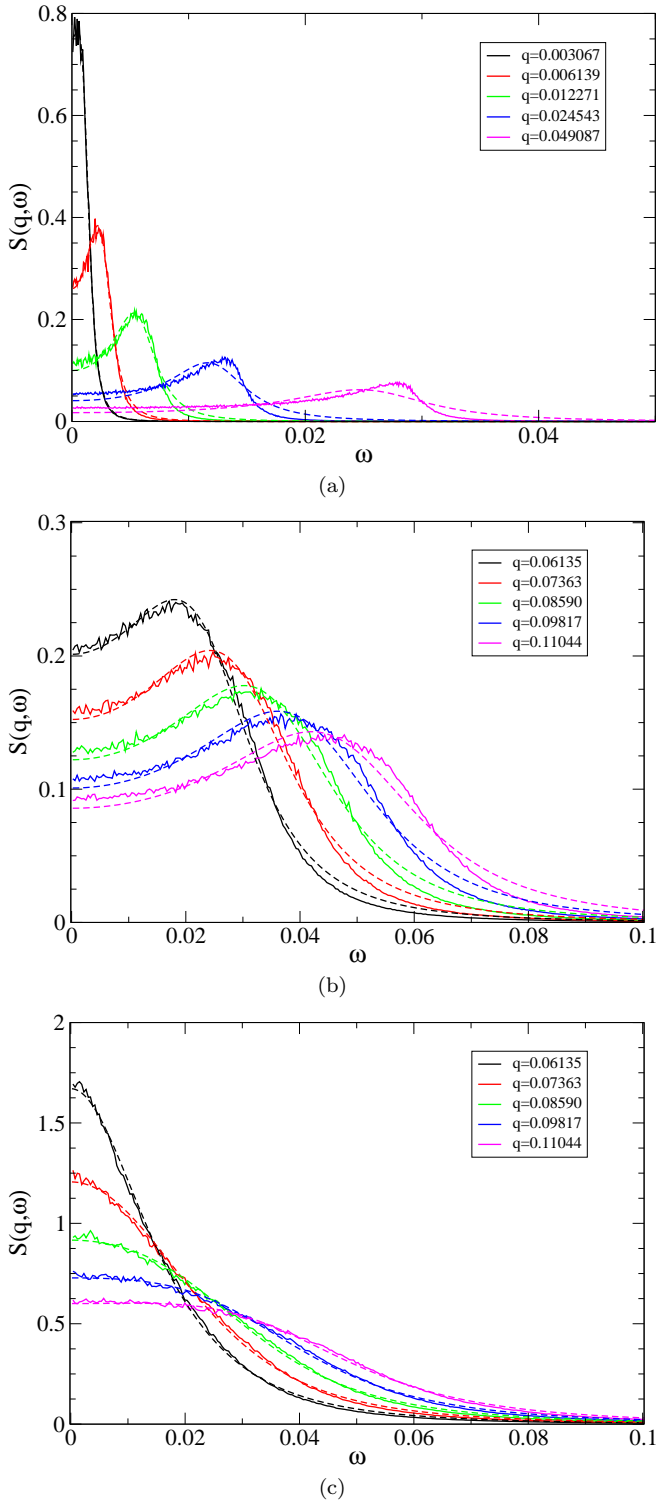

\begin{center}
\subfigure[]{\includegraphics[width=\columnwidth,clip]{Fig3a.eps}}
\subfigure[]{\includegraphics[width=\columnwidth,clip]{Fig3b.eps}}
\subfigure[]{\includegraphics[width=\columnwidth,clip]{Fig3c.eps}}
\caption{Power spectrum of energy fluctuations for a chain 
of $N=2048$ $C=1$ and (a) $\epsilon=0.1$ (b) 
$\epsilon=0.5$ and (c) $\epsilon=1.0$.
Dashed lines are best fit with Eq.(\ref{lorenz}).}
\label{sp}
\end{center}
\end{figure}
%%%%%%%%%%%%%%%%%%%%%%%%%%%%%%%%%%%%%%%%%%%%%%%%%%%%%%%%%%%%%%%%%%%%%%%%%%%%%%%%%%%%%%%%%%%

In a second series of simulations, we 
computed the dynamical structure factor, namely the square modulus of
temporal Fourier transform of the energy density on the lattice
\begin{equation}
e_n =\frac{\dot u_n^2}{2}  + \frac{u_n^2}{2}
+ \frac{u_n^4}{4} + \frac{C}{2}(u_{n+1} - u_n)^2  
\label{energy}
\end{equation}
\begin{equation}
e(q,t)  \;=\; \frac1N \,\sum_n \, e_n\exp(-iqn) \quad ,
\label{dens}
\end{equation}
which is defined as
\begin{equation}
S(q,\omega) \;=\;
\big\langle \big| e(q,\omega )\big|^2 \big\rangle  \quad .
\label{strutf}
\end{equation}
The square brackets denote an average over a set of independent
molecular--dynamics runs. By virtue of the periodic boundaries, the allowed
values of the  wavenumber $q$ are integer multiples of $2\pi/N$.

%Throughout the paper, temperature and energy will be 
%used interchageably because the heat capacity is always close 
%to one for the considered temperature range.  

The data in Fig.~\ref{sp} are representative of the numerical results.
At low enough temperatures we see a peak at finite-frequency which 
suggests some kind of oscillating response i.e. the propagation of 
damped temperature--waves. Upon increasing $\epsilon$ and/or decreasing $q$, the 
spectra display a central peak, akin to the one of an overdamped oscillator
which signals the onset of diffusive behavior. 

\begin{figure}
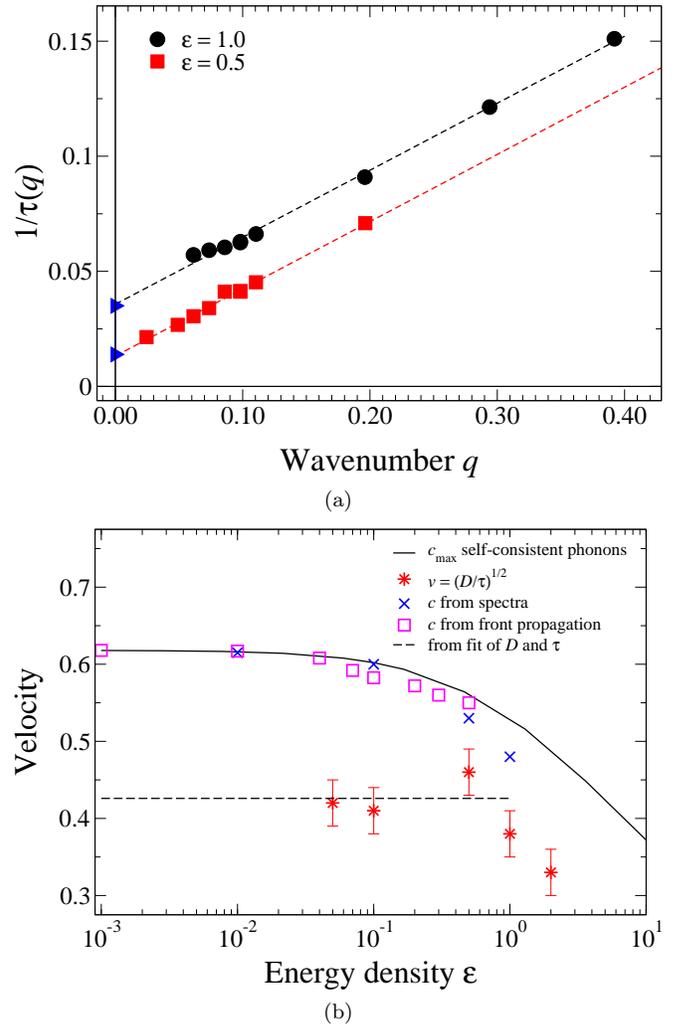

\begin{center}
\subfigure[]{\includegraphics[width=\columnwidth,clip]{Fig4a.eps}}
\subfigure[]{\includegraphics[width=\columnwidth,clip]{Fig4b.eps}}
\caption{(a) Dependence of $\tau(q)$ on the wavenumber for two
different energy densities. Lines are best fit with the law (\ref{fit}).
Triangles are the data for $\tau$ reported  Fig.~\ref{times}.
(b) Comparison of the relevant velocities emerging 
from linear response. The velocity $c$ defined from 
Eq.~(\ref{lorenz}) (dots) is very close to the maximum 
group velocity computed from the effective dispersion, 
Eq.~(\ref{eff}) (solid line). The thermal velocity $v$
is defined by Eq.~\eqref{v} (see text). Data are obtained 
for a chain of length  $N=2048$ with $C=1$.
}
\label{vel}
\end{center}
\end{figure}

For not too large $q$ the spectra are well fitted by a Lorentzian shape
\begin{equation}
S(q,\omega)=\frac{S_0}{(\omega^2 - \omega^2_0(q))^2 + ({\omega}/{\tau(q)})^2}
\quad .
\label{lorenz}
\end{equation}
We found that the dependence of the parameters $\omega_0(q)$ and
$\tau(q)$  from the wavenumber $q$ is  
\begin{equation}
\omega_0(q) \;=\; c|q| \quad, \qquad  
\frac{1}{\tau(q)} = \frac{1}{\tau} \left[1 + \left|\frac{q}{q_0}\right|\right]
\label{fit}
\end{equation}
where $c$ and $q_0$ are fitting parameters (see Fig.~\ref{vel}).

%%%%%%%%%%%%%%%%%%%%%%%%%%%%%%%%%%%%%%%%%%%%%%%%%%%%%%%%%%%%%%%%%%%%%%%%%%%%%%
%%%%%%%%%%%%%%%%%%%%%%%%%%%%%%%%%%%%%%%%%%%%%%%%%%%%%%%%%%%%%%%%%%%%%%%%%%%%%%

The observed form of the line-shape, Eq.~(\ref{lorenz}),  is fully
consistent with what expected from Eq.~(\ref{teleg}). However, the origin
of the dependence of $\tau(q)$ cannot be accounted for by the simplified
kernel (\ref{catt}) and is presumably  a signature of spatial
correlations. Nonetheless, we note that the value $\tau$ obtained by
extrapolating at $q\to0$ in the second of  Eqs.~(\ref{fit}) is in very
good agreement with the value measured from flux spectra (see again 
Fig~\ref{times}).  Another
important quantitative difference is in the characteristic velocities. In
Fig.~\ref{vel}b we compare the velocity $c$ from the fitting (\ref{fit})
with $v$ as given by definition (\ref{v}). Since we found above that $D$
and $\tau$ are both proportional to (roughly) the same  power-law
$\epsilon^{-\beta}$, in  Fig.~\ref{vel} we report also the ratio of the 
corresponding proportionality constants. The measured values of $v$ in 
Fig.~(\ref{vel}) show that $v\lesssim c$ in the considered range. 

We may surmise that the velocity $c$ should be related to the 
group velocity of the harmonic waves. To account for 
finite-temperature effects we have computed the renormalized 
dispersion relation in the self--consistent phonon approximation 
~\cite{dauxois},
\begin{equation}
\tilde\Omega^2(q) \;=\; \omega^2(T) + 2C(1 -\cos q)\quad.
\label{eff}
\end{equation}
where 
\begin{equation}
\omega^2(T) = 1 + \frac{\langle u^4\rangle}{\langle u^2\rangle} 
\end{equation}
The second and fourth momenta of the displacements have been evaluated
numerically. The data for $c$ are very close to the maximum group velocity
$c_{max}$, namely the largest value of 
$d\tilde\Omega(q)/dq$ as computed from formula \eqref{eff}, see Fig.~\ref{vel}.

We conclude that, although the telegraph equation~(\ref{teleg})
accounts for the line-shape of the energy correlators, there
are, at least in the considered times and length ranges, some 
quantitative deviations. 
The fact that $v$ and $c$ are different can be partly understood 
by noting that $v$ is associated with diffusive processes and results
from interaction of all possible lattice waves. The maximum group
velocity $c$ will thus be an upper bound to $v$, but 
the two need not be equal \cite{preziosi}. It is interesting to mention that 
differences in the measured velocities were previously reported
also in a model of hard--point particles ~\cite{Delfini07a} .

%-------------------------------------------------------------------------------------
\section{Heat pulse propagation}
\label{sec:pulse}
%%-------------------------------------------------------------------------------------

%%%%%%%%%%%%%%%%%%%%%%%%%%%%%%%%%%%%%%%%%%%%%%%%%%%%%%%%%%%%%%%%%%%%%%%%%%%%%%%%%%%%%%
%\begin{widetext}
\begin{figure*}
\begin{center}
\subfigure[]{\includegraphics[width=\columnwidth,clip]{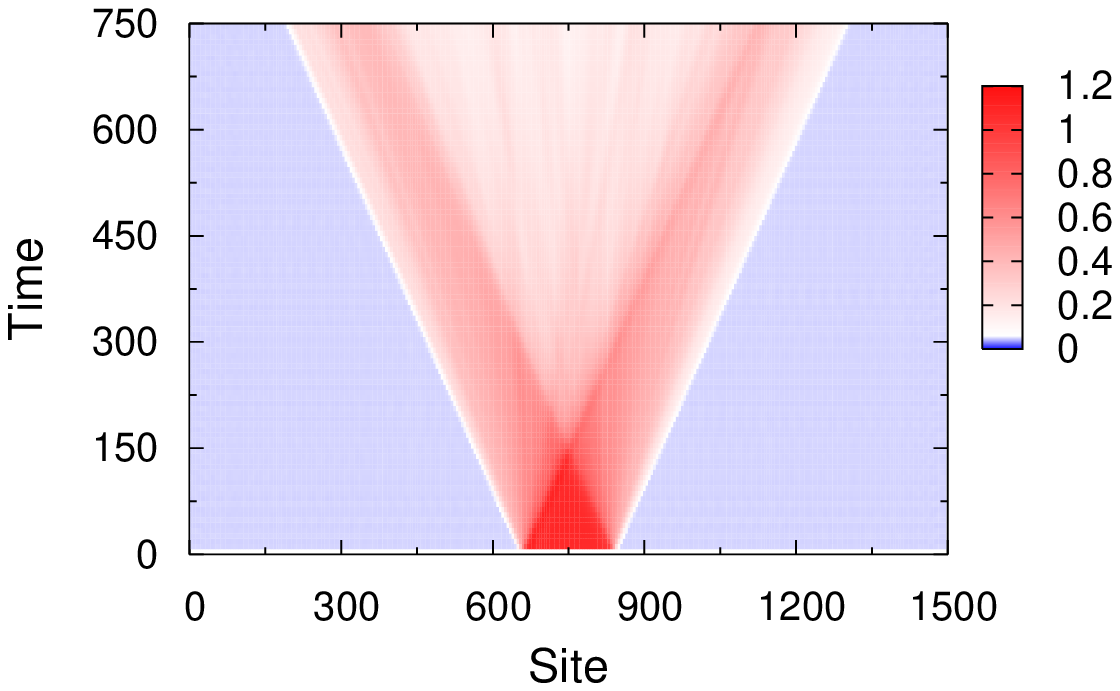}}
\subfigure[]{\includegraphics[width=\columnwidth,clip]{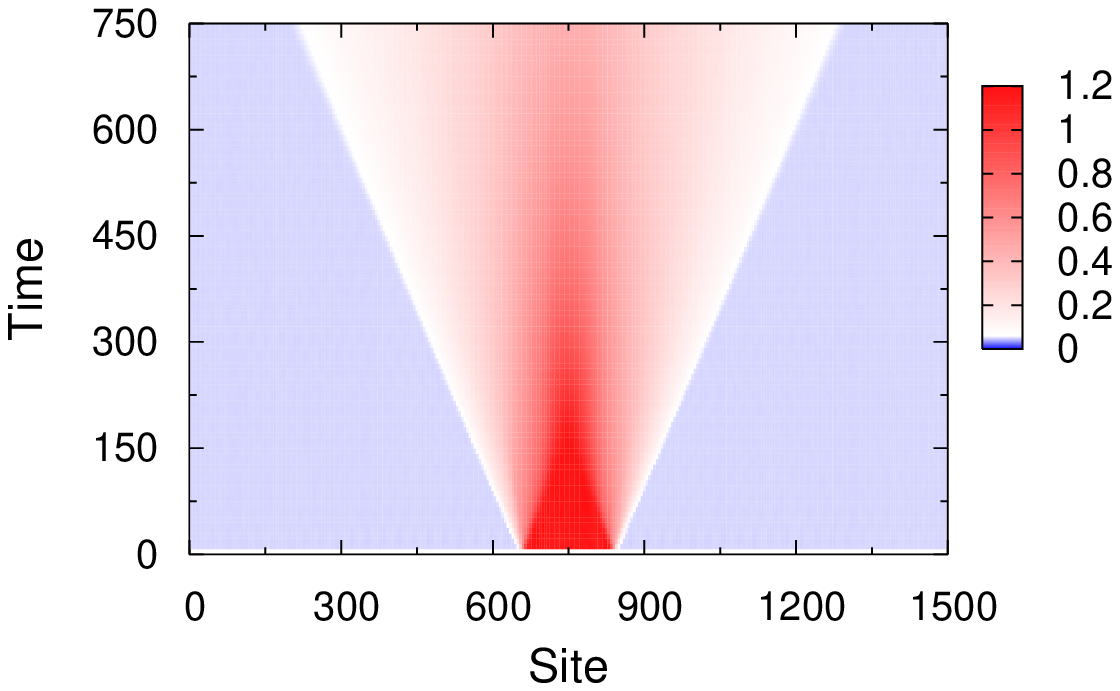}}
\caption{Spatio-temporal density maps of the normalized temperature profiles $T(n,t)/T_{0}$
of two heat pulses propagating in chains at temperatures $T_{b}=0.001$ (a) and 
$T_{b}=0.04$ (b). Other parameters are: $T_{0}=20 \times T_{b}$, $C=1$, $N=1500$.}
\label{f:TprofMAPS}
\end{center}
\end{figure*}
%\end{widetext}
%%%%%%%%%%%%%%%%%%%%%%%%%%%%%%%%%%%%%%%%%%%%%%%%%%%%%%%%%%%%%%%%%%%%%%%%%%%%%%%%%%%%%%

The linear response results reported so far suggest that the propagation of heat waves 
could be described at the macroscopic level, at least for small enough energies, 
through  the telegraph equation~(\ref{teleg}). In order to test this conjecture, we performed nonequilibrium
numerical simulations where a heat pulse of temperature $T_{0}$ is excited in a small region of width $\Delta N$
within  a chain otherwise at equilibrium at the {\em bulk} temperature $T_{b} < T_{0}$ and observed as it evolves.
A typical pulse experiment proceeds as follows. First the chain is let evolve at a fixed energy 
$\epsilon = T_{b}$ for a long enough equilibration time. Subsequently, the portion $[(N-\Delta N)/2,(N+\Delta N)/2]$
is put in contact with a Langevin thermostat at the temperature $T_{0}$, while the rest of the chain is kept 
frozen in its equilibrium configuration. When the heated portion has reached thermal equilibrium with the thermostat,
the latter is removed and the equations of motion of the whole chain are integrated at constant energy.
For the pulse propagation experiments we used the symplectic ``Position
Extended Forest-Ruth Like'' (PEFRL) algorithm of Omelyan et al.~\cite{PEFRL} and a velocity Verlet algorithm 
for the microcanonical and Langevin integrations, respectively. Furthermore, 
since we used free-ends boundary conditions, we took care to chose the integration time so as to avoid 
the propagating pulse to be reflected at the chain edges. In order to obtain highly accurate results, we
averaged the time-dependent temperature profiles over a large ensemble of ${\mathcal N}$ independent realizations of the
equilibrium configuration of both the bulk and the excited regions of the chain. All results reported in the following 
were obtained with ${\mathcal N}= 2 \times 10^4$ and $\Delta N = 90$, except where explicitly indicated otherwise.

Typical spatio-temporal portraits of the temperature profiles describing the evolution of a heat pulse
are reported in Fig.~\ref{f:TprofMAPS} for two different values of the bulk temperature. For small values of $T_{b}$, 
the heat excitation gives rise to two symmetrical pulses, traveling in opposite  directions at a constant speed.
Such speed is extremely close to the maximum group velocity of the linear modes, $c_{max}=0.618$. As the bulk 
temperature is increased, the correlation time $\tau$ is expected to decrease thus reducing the window of
ballistic propagation. In fact, a diffusion-like
evolution of the pulse is observed already at $T_{b}=0.04$ (Fig.~\ref{f:TprofMAPS}b). However, it can be clearly 
appreciated that the bulk temperature is still not sufficiently large for the pulse to spread homogeneously at 
all heights. At a closer inspection, it is not difficult to realize that the the front baseline still clearly 
moves at a constant speed close to $c_{max}$. 

\begin{figure*}
\begin{center}
\subfigure[]{\includegraphics[width=\columnwidth,clip]{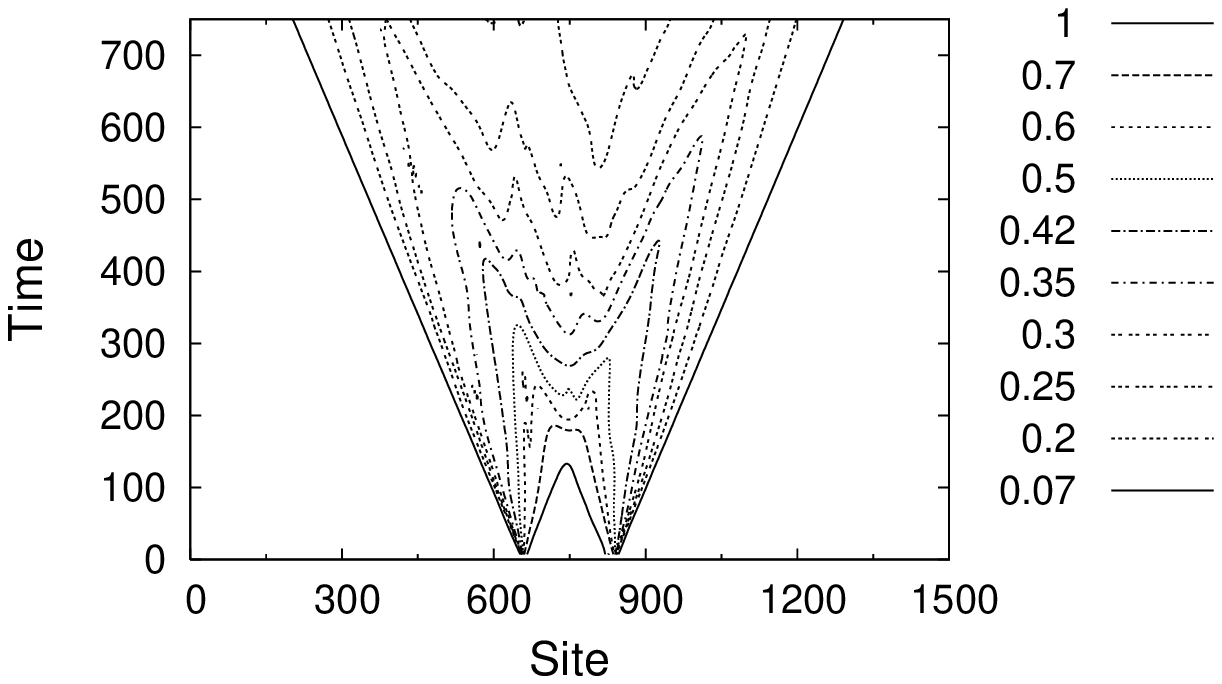}}
\subfigure[]{\includegraphics[width=\columnwidth,clip]{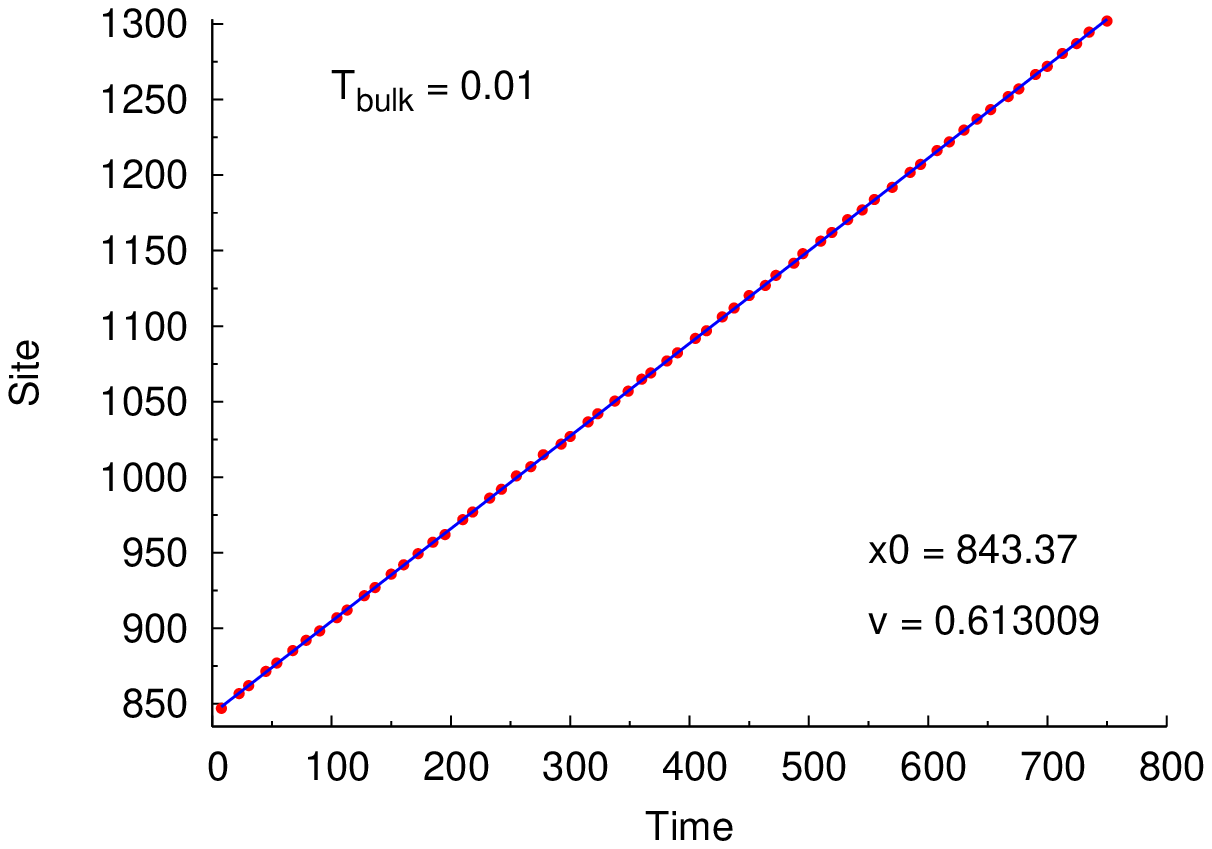}}
\subfigure[]{\includegraphics[width=\columnwidth,clip]{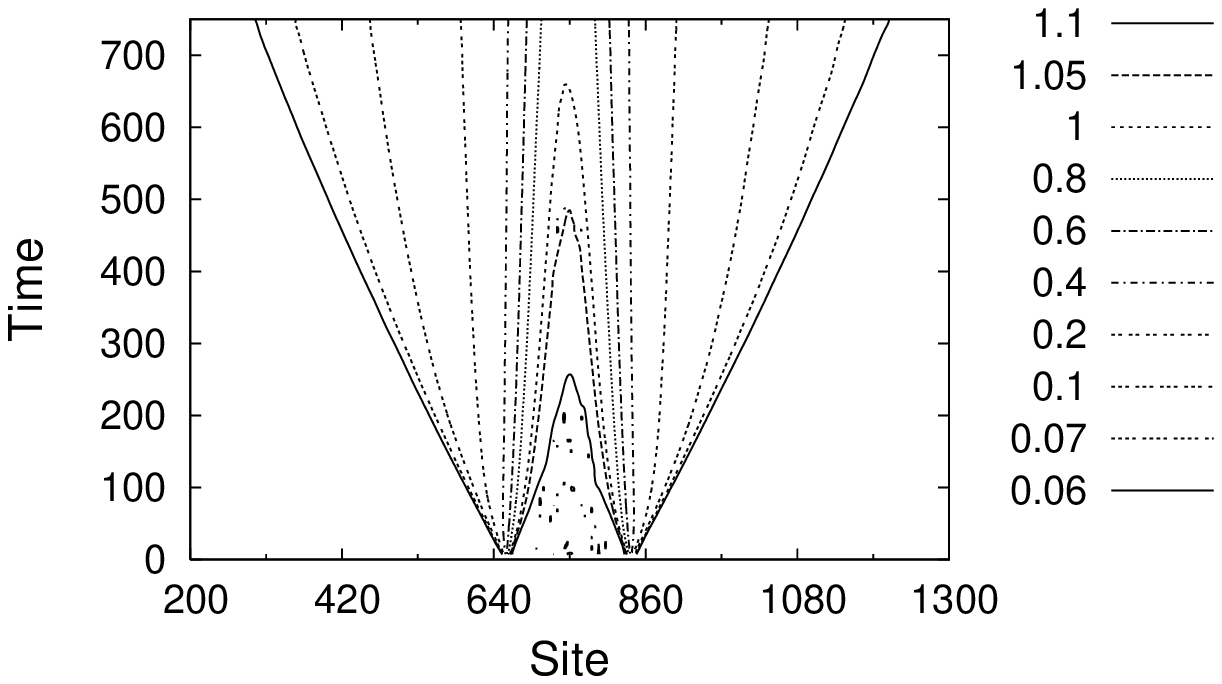}}
\subfigure[]{\includegraphics[width=\columnwidth,clip]{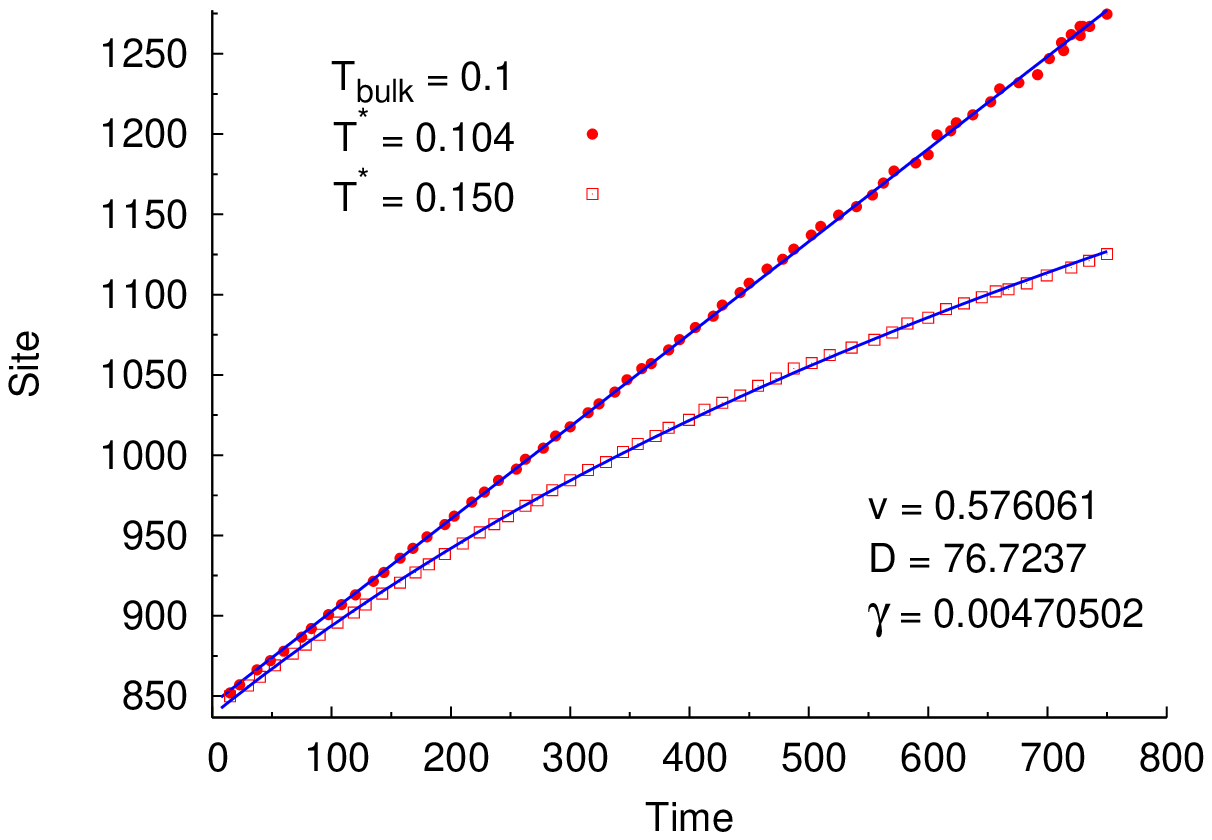}}
\subfigure[]{\includegraphics[width=\columnwidth,clip]{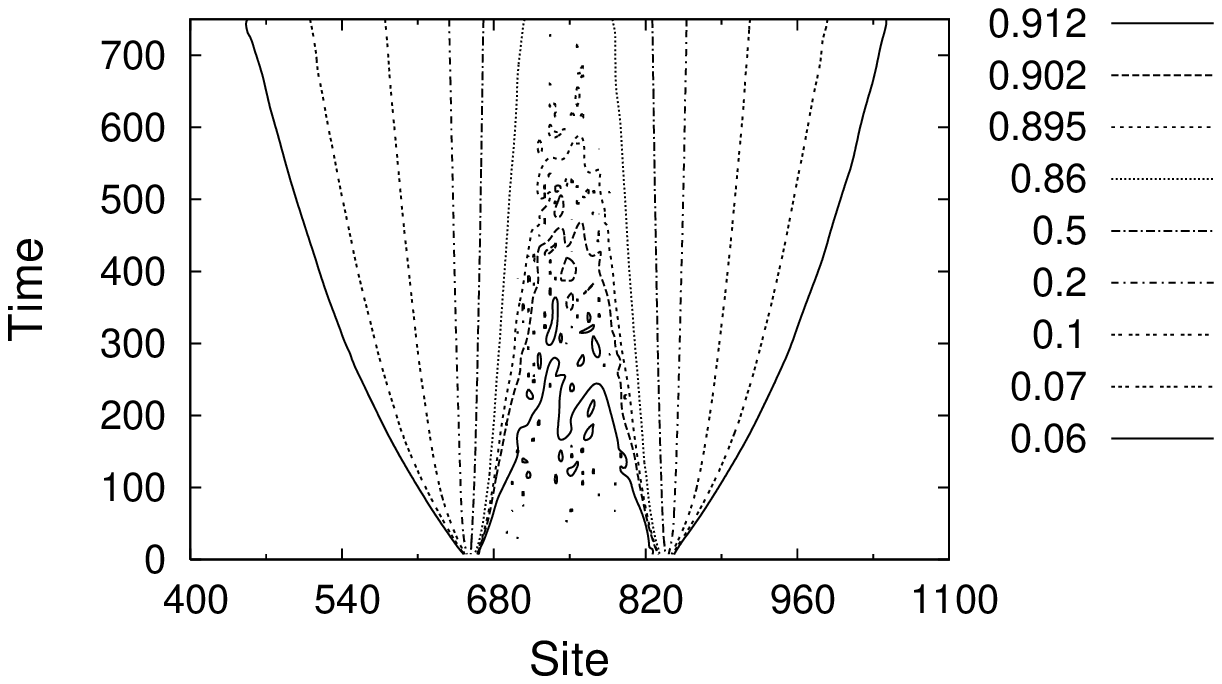}}
\subfigure[]{\includegraphics[width=\columnwidth,clip]{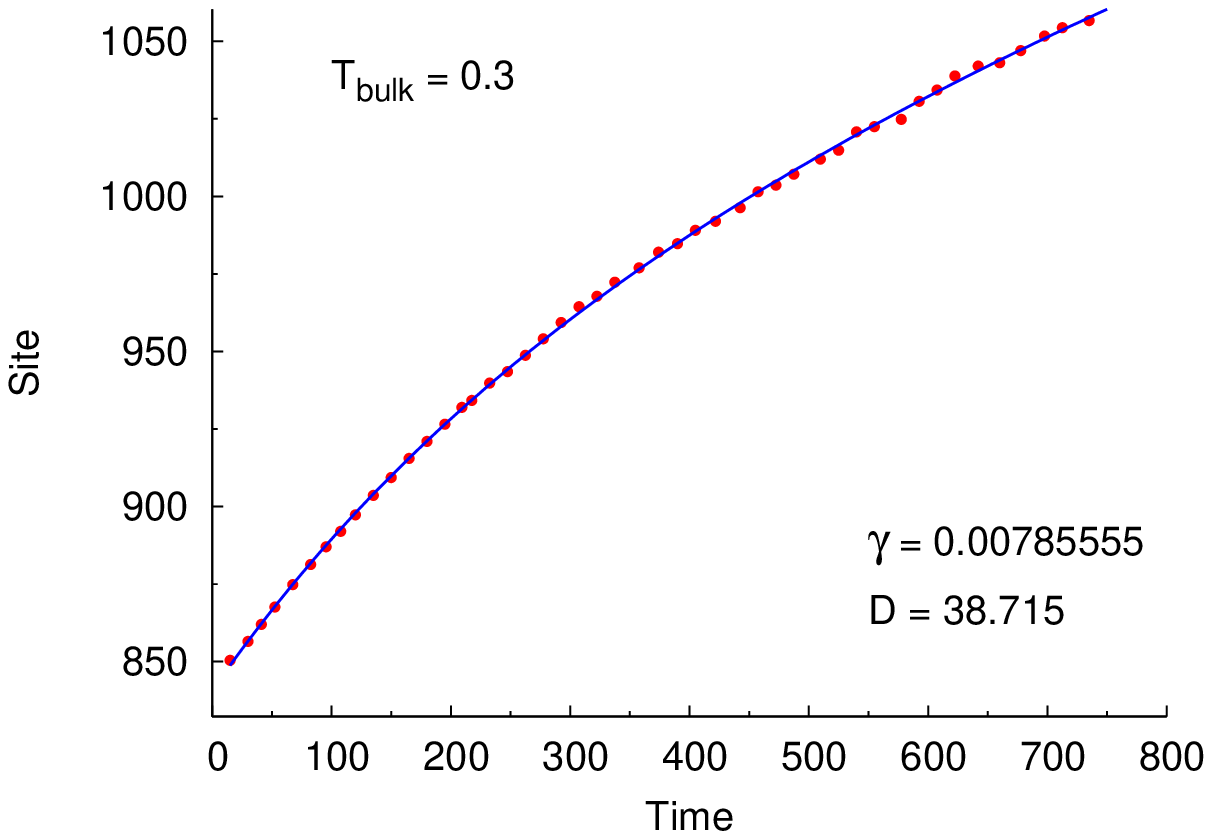}}
\caption{Analysis of heat pulse propagation. Spatio-temporal 
contour maps of the normalized temperature profiles $T(n,t)/T_{0}$ (left panels) and time variation of the front 
baseline position at a given temperature height: $T^\ast=0.0105$ (b), $T^\ast=0.104$ and 
$T^\ast=0.15$ (d), $T^\ast=0.35$ (f). Parameters are.
$C=1$, $T_{b}=0.01$ (a) and (b), $T_{b}=0.1$  (c) and (d),
$T_{b}=0.3$  (e) and (f), $T_{0}=20 \times T_{b}$, $C=1$, $N=1500$. The straight lines  in panels (b) and (c) are
linear fits, while the other solid lines are fits with formula~\eqref{e:x2prw}.}
\label{f:Tprof2D}
\end{center}
\end{figure*}
%%%%%%%%%%%%%%%%%%%%%%%%%%%%%%%%%%%%%%%%%%%%%%%%%%%%%%%%%%%%%%%%%%%%%%%%%%%%%%%%%%%%%%

Stated more precisely, our conjecture implies that, if the telegraph equation holds, then the motion of the 
heat pulse front should carry the information on the microscopic parameters that enter the continuum description,
namely the speed $v$ and the correlation time $\tau$ (see again equation~(\ref{teleg})).
Let us imagine to {\em cut} the time-dependent temperature profiles that describe the chain dynamics 
after excitation of  the heat pulse at a given temperature $T^\ast$, with $T_{b} < T^\ast \ll T_{0}$.
Then, one may extract useful information by tracking the sites occupied by the pulse front at the 
height $T^\ast$ during its evolution. In other words, one may study the trajectories defined implicitly as
\be
\label{trajx2av}
T(x(t),t) = T^\ast
\ee
where $T(n,t)  = \langle \dot{u}^2_{n}(t) \rangle$ is the temperature field , the 
average being computed over an ensemble of ${\mathcal N}$ independent trajectories. 
If $T^\ast$ is sufficiently close
to the bulk temperature $T_{b}$, the linear response results should hold, and the
persistent random walk should be recovered, namely 
\be
\label{e:x2prw}
x(t) - x_{0} \simeq \sqrt{\frac{2v^2}{\gamma} 
      \left[
         t - \frac{1}{\gamma} \left(1 - e^{-\gamma t} \right) 
      \right]}
\ee
The analysis summarized in Fig.~\ref{f:Tprof2D} proves the validity of our inference. 
Close to the background temperature, the propagation crosses-over from ballistic 
to diffusive on a time scale that is well predicted by the equilibrium
simulations (see Fig.~\ref{times}). For $T_{b}=0.01$, the equilibrium 
prediction would be a time scale of the order $1/\gamma \approx 2 \times 10^4$.
Correspondingly, on our observation window we observe purely ballistic propagation
(upper right panel  of Fig.~\ref{f:Tprof2D}). Conversely, 
for $T_{b}=0.3$ the equilibrium relaxation time is about  150. Accordingly, we indeed
observe  a crossover to diffusion within the time span of our simulations.
From the  fit we get $1/\gamma = 127$, in good agreement with the equilibrium 
prediction (lower right panel  of Fig.~\ref{f:Tprof2D}). 
At intermediate  temperatures, an increase of the  {\em cut} temperature height
in the vicinity of $T_{b}$ causes the ballistic/diffusive cross-over to
occur on shorter time scales, as expected since the relaxation time approaches the 
observation time. In fact, this can be seen as the very definition of
the {\em intermediate} time scale.  This is clearly illustrated in the middle right panel 
of Fig.~\ref{f:Tprof2D} ($T_{b}=0.1$), where the transition becomes visible in the simulation 
time window upon raising the cut temperature from $T^\ast=0.104$ to $T^\ast=0.15$.

As a further check of consistency, we have examined the time-variations of the 
temperature field second moment $\mu_{2}(t)$, defined as
\be
\mu_{2}(t) = \sum_{n} (n - \langle n \rangle)^2 z_{n} 
\label{e:mu2}
\ee
where $z_{n} = \dot{u}^2/\sum_{m} \dot{u}^2_{m}$ and $\langle n \rangle  = \sum_{n} n z_{n}$.
However, straight calculation of the second moment from its definition requires extremely accurate 
averages, due to the strong amplification of the fluctuations in the bulk regions away from 
$ \langle n \rangle$. For this reason, in order to monitor $\mu_{2}$ over time, 
two copies of the system were evolved starting from identical initial conditions but for the 
small region around the center of the chain, where the heat pulse is generated.
An accurate estimate of the pulse spread for finite values of $T_{b}$ 
can then be obtained from the second moment of the difference between the profiles
of the two system clones.

%%%%%%%%%%%%%%%%%%%%%%%%%%%%%%%%%%%%%%%%%%%%%%%%%%%%%%%%%%%%%%%%%%%%%%%%%%%%%%%%%%%%%%
\begin{figure}[ht!]
\begin{center}
\includegraphics[width=\columnwidth,clip]{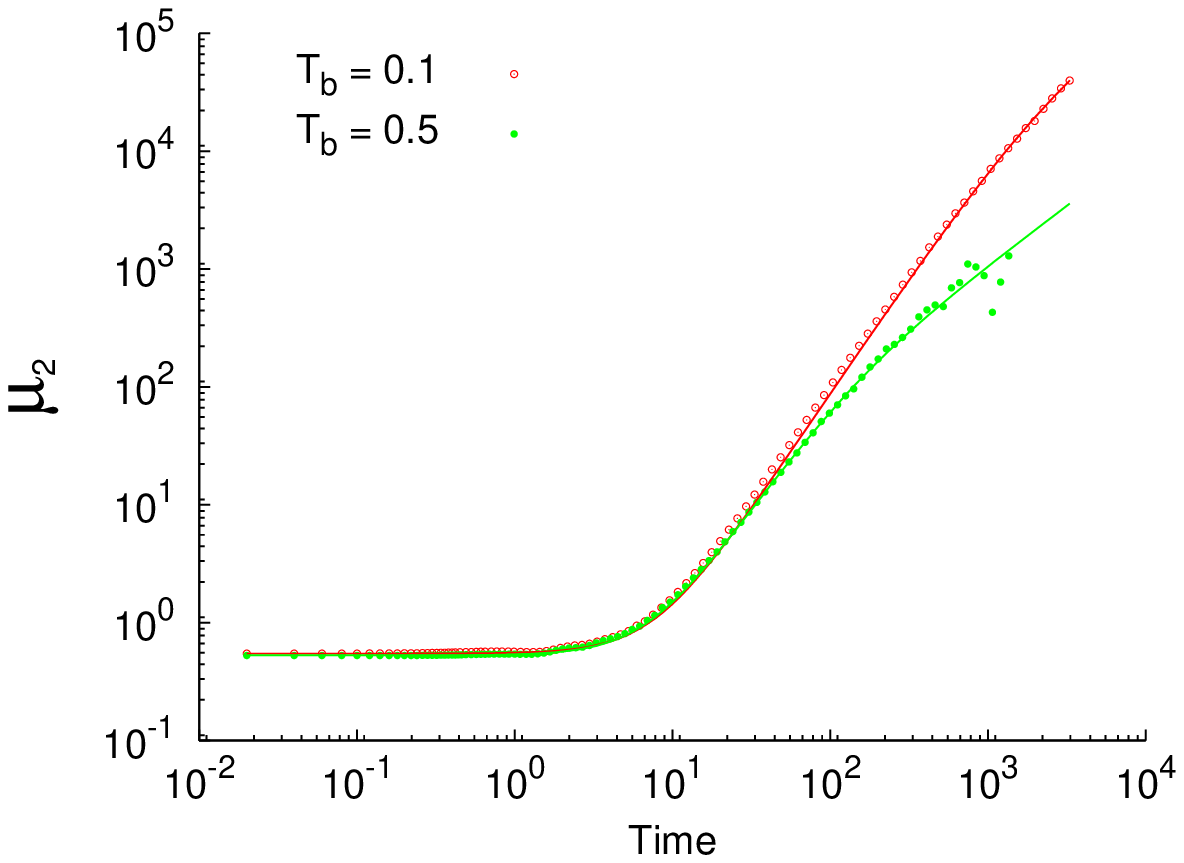}  
\caption{Time evolution of the second moment of the temperature field 
for two values of the bulk temperature, $T_{b}=0.1$ ($T_{0}=0.5$) and
$T_{b}=0.5$ ($T_{0}=1.0$) (symbols) and two-parameter fits with expression~\eqref{e:mu2tel}.
The initial conditions $\mu_{2}(0)$ have been fixed at the corresponding 
values extracted from the numerics.
Other parameters are: $C=1$, $N=5000$, $\Delta N= 10$.}
\label{f:heights1}
\end{center}
\end{figure}
%%%%%%%%%%%%%%%%%%%%%%%%%%%%%%%%%%%%%%%%%%%%%%%%%%%%%%%%%%%%%%%%%%%%%%%%%%%%%%%%%%%%%%

The result of such procedure is  shown in Fig.~\ref{f:heights1} for two values
of the bulk temperature in the case of a narrow excitation. As it shows, the heat pulse spreads
according to the prescription of the telegraph equation, that is
\be
\label{e:mu2tel}
\mu_{2}(t)  = \mu_{2}(0) + \frac{2v^2}{\gamma} 
      \left[
         t - \frac{1}{\gamma} \left(1 - e^{-\gamma t} \right) 
      \right]
\ee
The time scales extracted from the fits are $\tau = 917.5$ ($T_{b}= 0.1$) and 
$\tau = 53.9$ ($T_{b}= 0.5$) . These
figures are in good agreement with the equilibrium relaxation times that we found
by fitting the results of microcanonical simulations (data shown in Fig.~\ref{times}), 
$\tau = 760.8$ ($T_{b}= 0.1$) and  $\tau = 76.2$ ($T_{b}= 0.5$).

The best-fit  values of the velocities $v$ prove rather insensitive to  the bulk
temperature. We find $v = 0.095$  ($T_{b}= 0.1$) and $v = 0.1$  ($T_{b}= 0.5$).
Interestingly, these values are smaller than all estimates shown in
Fig.~\ref{vel}. This is likely to reflect the temperature-dependence  of the
relaxation times at different heights within the pulse. For a finite temperature
disturbance, the fronts do not spread at the same rate, causing the hotter
portions to lag behind the advancing baseline. Overall, this should reflect in a
lower value of the pulse velocity during its first ballistic stage.

\section{Discussion}

In this report we have investigated the relaxation of temperature fluctuations
in a discrete nonlinear chain. At low temperatures the relaxation time $\tau$ 
of the energy current must be taken into account, leading to correction to the
standard diffusive behavior. Starting from numerical calculation of the
response function and to the simplest level of approximation, we obtained the
telegraph equation, Eq.~(\ref{teleg}), and estimated the  temperature
dependence of its parameters. 

The comparison between the linear-response prediction and the nonequilibrium
simulations reveals that the telegraph equation provides a reliable {\em macroscopic}
interpretative framework for quantifying the front propagation. 
In particular, we have shown that the second moment of the temperature field 
displays the ballistic/diffusive cross-over on a time scale in accordance with 
the correlation times of energy fluctuations extracted from equilibrium simulations. 
The same is true for the front propagation at temperatures close to the 
background. 
It might be surmised that this type of macroscopic description should apply
to all models displaying normal energy transport, such as chains of coupled
rotors or other one-dimensional models with pinning potentials~\cite{LLP03},
or three-dimensional systems~\cite{Volz96}.

By following the propagation at higher temperatures, the front appears to smear out in
the  course of time. It is likely that this effect  could be captured by a kernel of the
Jeffreys type, as shown by Maruyama and co-workers~\cite{maruyama}. Another possible 
improvement would be to include spatial memory effects,  by allowing for a space-dependence 
memory dependence in the kernel $K$, Eq.(\ref{fourier}). This is especially important if one
wishes to describe systems with anomalous transport properties~\cite{LLP03}. 
Indeed, in this case heat propagation is 
quantitatively described by a Levy walk process~\cite{denisov,Delfini07a},  
which is precisely the generalization of the persistent random walk 
for of a memory decaying as a power law.
As a consequence, the macroscopic equation generalizing the telegraph 
equation should involve mixed spatio-temporal fractional derivatives ~\cite{Sokolov03}. 
Possible nonlinear heat-wave propagation could also be taken into account by
including higher--order powers of the local gradient. Further work along these
lines is in progress. 

Finally, we remark that our results confirm that wave-like transport of energy 
is associated with the presence of optical branches in the linear spectrum.
This could be of interest for heat transport in
single-walled carbon nanotubes, where a significant contribution of optical phonons 
to wave-like conduction has been reported~\cite{maruyama}. 
In this respect one may conjecture that our simplified model may provide
an effective description of more complex quasi-one-dimensional nanostructures.

\section*{Acknowledgements}

We thank Carlos Mejia-Monasterio, Luca Delfini, Marc Weber and Paolo De Los Rios 
for useful discussions. 
SL acknowledges support of the Fonds National Suisse de la 
Recherche Scientifique (SNF), through the individual grant 
\textit{Localization and transport in nonlinear systems}. 

%\bibliography{teleg}

\begin{thebibliography}{26}
\expandafter\ifx\csname natexlab\endcsname\relax\def\natexlab#1{#1}\fi
\expandafter\ifx\csname bibnamefont\endcsname\relax
  \def\bibnamefont#1{#1}\fi
\expandafter\ifx\csname bibfnamefont\endcsname\relax
  \def\bibfnamefont#1{#1}\fi
\expandafter\ifx\csname citenamefont\endcsname\relax
  \def\citenamefont#1{#1}\fi
\expandafter\ifx\csname url\endcsname\relax
  \def\url#1{\texttt{#1}}\fi
\expandafter\ifx\csname urlprefix\endcsname\relax\def\urlprefix{URL }\fi
\providecommand{\bibinfo}[2]{#2}
\providecommand{\eprint}[2][]{\url{#2}}

\bibitem[{\citenamefont{Joseph and Preziosi}(1989)}]{preziosi}
\bibinfo{author}{\bibfnamefont{D.~D.} \bibnamefont{Joseph}} \bibnamefont{and}
  \bibinfo{author}{\bibfnamefont{L.}~\bibnamefont{Preziosi}},
  \bibinfo{journal}{Rev. Mod. Phys.} \textbf{\bibinfo{volume}{61}},
  \bibinfo{pages}{41} (\bibinfo{year}{1989}).

\bibitem[{\citenamefont{Lepri et~al.}(2003)\citenamefont{Lepri, Livi, and
  Politi}}]{LLP03}
\bibinfo{author}{\bibfnamefont{S.}~\bibnamefont{Lepri}},
  \bibinfo{author}{\bibfnamefont{R.}~\bibnamefont{Livi}}, \bibnamefont{and}
  \bibinfo{author}{\bibfnamefont{A.}~\bibnamefont{Politi}},
  \bibinfo{journal}{Phys. Rep.} \textbf{\bibinfo{volume}{377}},
  \bibinfo{pages}{1} (\bibinfo{year}{2003}).

\bibitem[{\citenamefont{Cipriani et~al.}(2005)\citenamefont{Cipriani, Denisov,
  and Politi}}]{denisov}
\bibinfo{author}{\bibfnamefont{P.}~\bibnamefont{Cipriani}},
  \bibinfo{author}{\bibfnamefont{S.}~\bibnamefont{Denisov}}, \bibnamefont{and}
  \bibinfo{author}{\bibfnamefont{A.}~\bibnamefont{Politi}},
  \bibinfo{journal}{Phys. Rev. Lett.} \textbf{\bibinfo{volume}{94}},
  \bibinfo{eid}{244301} (\bibinfo{year}{2005}).

\bibitem[{\citenamefont{Zhao}(2006)}]{zhao}
\bibinfo{author}{\bibfnamefont{H.}~\bibnamefont{Zhao}}, \bibinfo{journal}{Phys.
  Rev. Lett.} \textbf{\bibinfo{volume}{96}}, \bibinfo{eid}{140602}
  (\bibinfo{year}{2006}).

\bibitem[{\citenamefont{Delfini et~al.}(2007)\citenamefont{Delfini, Denisov,
  Lepri, Livi, Mohanty, and Politi}}]{Delfini07a}
\bibinfo{author}{\bibfnamefont{L.}~\bibnamefont{Delfini}},
  \bibinfo{author}{\bibfnamefont{S.}~\bibnamefont{Denisov}},
  \bibinfo{author}{\bibfnamefont{S.}~\bibnamefont{Lepri}},
  \bibinfo{author}{\bibfnamefont{R.}~\bibnamefont{Livi}},
  \bibinfo{author}{\bibfnamefont{P.~K.} \bibnamefont{Mohanty}},
  \bibnamefont{and} \bibinfo{author}{\bibfnamefont{A.}~\bibnamefont{Politi}},
  \bibinfo{journal}{Europ. Phys. J. - Special Topics}
  \textbf{\bibinfo{volume}{146}}, \bibinfo{pages}{21} (\bibinfo{year}{2007}).

\bibitem[{\citenamefont{Helfand}(1960)}]{Helfand60}
\bibinfo{author}{\bibfnamefont{E.}~\bibnamefont{Helfand}},
  \bibinfo{journal}{Phys. Rev.} \textbf{\bibinfo{volume}{119}},
  \bibinfo{pages}{1} (\bibinfo{year}{1960}).

\bibitem[{\citenamefont{Cahill et~al.}(2003)\citenamefont{Cahill, Ford,
  Goodson, Mahan, Majumdar, Maris, Merlin, and Phillpot}}]{cahill}
\bibinfo{author}{\bibfnamefont{D.~G.} \bibnamefont{Cahill}},
  \bibinfo{author}{\bibfnamefont{W.~K.} \bibnamefont{Ford}},
  \bibinfo{author}{\bibfnamefont{K.~E.} \bibnamefont{Goodson}},
  \bibinfo{author}{\bibfnamefont{G.~D.} \bibnamefont{Mahan}},
  \bibinfo{author}{\bibfnamefont{A.}~\bibnamefont{Majumdar}},
  \bibinfo{author}{\bibfnamefont{H.~J.} \bibnamefont{Maris}},
  \bibinfo{author}{\bibfnamefont{R.}~\bibnamefont{Merlin}}, \bibnamefont{and}
  \bibinfo{author}{\bibfnamefont{S.~R.} \bibnamefont{Phillpot}},
  \bibinfo{journal}{J. App. Phys.} \textbf{\bibinfo{volume}{93}},
  \bibinfo{pages}{793} (\bibinfo{year}{2003}).

\bibitem[{\citenamefont{Hone et~al.}(1999)\citenamefont{Hone, Whitney, Piskoti,
  and Zettl}}]{hone}
\bibinfo{author}{\bibfnamefont{J.}~\bibnamefont{Hone}},
  \bibinfo{author}{\bibfnamefont{M.}~\bibnamefont{Whitney}},
  \bibinfo{author}{\bibfnamefont{C.}~\bibnamefont{Piskoti}}, \bibnamefont{and}
  \bibinfo{author}{\bibfnamefont{A.}~\bibnamefont{Zettl}},
  \bibinfo{journal}{Phys. Rev. B} \textbf{\bibinfo{volume}{59}},
  \bibinfo{pages}{R2514} (\bibinfo{year}{1999}).

\bibitem[{\citenamefont{Yu et~al.}(2005)\citenamefont{Yu, Shi, Yao, Li, and
  Majumdar}}]{Yu05}
\bibinfo{author}{\bibfnamefont{C.}~\bibnamefont{Yu}},
  \bibinfo{author}{\bibfnamefont{L.}~\bibnamefont{Shi}},
  \bibinfo{author}{\bibfnamefont{Z.}~\bibnamefont{Yao}},
  \bibinfo{author}{\bibfnamefont{D.}~\bibnamefont{Li}}, \bibnamefont{and}
  \bibinfo{author}{\bibfnamefont{A.}~\bibnamefont{Majumdar}},
  \bibinfo{journal}{Nano Letters} \textbf{\bibinfo{volume}{5}},
  \bibinfo{pages}{1842} (\bibinfo{year}{2005}).

\bibitem[{\citenamefont{Osman and Srivastava}(2005)}]{osman}
\bibinfo{author}{\bibfnamefont{M.~A.} \bibnamefont{Osman}} \bibnamefont{and}
  \bibinfo{author}{\bibfnamefont{D.}~\bibnamefont{Srivastava}},
  \bibinfo{journal}{Phys. Rev. B} \textbf{\bibinfo{volume}{72}},
  \bibinfo{eid}{125413} (\bibinfo{year}{2005}).

\bibitem[{\citenamefont{Chester}(1963)}]{Chester}
\bibinfo{author}{\bibfnamefont{M.}~\bibnamefont{Chester}},
  \bibinfo{journal}{Phys. Rev.} \textbf{\bibinfo{volume}{131}},
  \bibinfo{pages}{2013} (\bibinfo{year}{1963}).

\bibitem[{\citenamefont{Shiomi and Maruyama}(2006)}]{maruyama}
\bibinfo{author}{\bibfnamefont{J.}~\bibnamefont{Shiomi}} \bibnamefont{and}
  \bibinfo{author}{\bibfnamefont{S.}~\bibnamefont{Maruyama}},
  \bibinfo{journal}{Phys. Rev. B} \textbf{\bibinfo{volume}{73}},
  \bibinfo{eid}{205420} (\bibinfo{year}{2006}).

\bibitem[{\citenamefont{Chang et~al.}(2008)\citenamefont{Chang, Okawa, Garcia,
  Majumdar, and Zettl}}]{Chang08}
\bibinfo{author}{\bibfnamefont{C.~W.} \bibnamefont{Chang}},
  \bibinfo{author}{\bibfnamefont{D.}~\bibnamefont{Okawa}},
  \bibinfo{author}{\bibfnamefont{H.}~\bibnamefont{Garcia}},
  \bibinfo{author}{\bibfnamefont{A.}~\bibnamefont{Majumdar}}, \bibnamefont{and}
  \bibinfo{author}{\bibfnamefont{A.}~\bibnamefont{Zettl}},
  \bibinfo{journal}{Phys. Rev. Lett.} \textbf{\bibinfo{volume}{101}},
  \bibinfo{pages}{075903} (\bibinfo{year}{2008}).

\bibitem[{\citenamefont{Hu et~al.}(1998)\citenamefont{Hu, Li, and Zhao}}]{Hu98}
\bibinfo{author}{\bibfnamefont{B.}~\bibnamefont{Hu}},
  \bibinfo{author}{\bibfnamefont{B.}~\bibnamefont{Li}}, \bibnamefont{and}
  \bibinfo{author}{\bibfnamefont{H.}~\bibnamefont{Zhao}},
  \bibinfo{journal}{Phys. Rev. E} \textbf{\bibinfo{volume}{57}},
  \bibinfo{pages}{2992} (\bibinfo{year}{1998}).

\bibitem[{\citenamefont{Aoki and Kusnezov}(2000)}]{aoki}
\bibinfo{author}{\bibfnamefont{K.}~\bibnamefont{Aoki}} \bibnamefont{and}
  \bibinfo{author}{\bibfnamefont{D.}~\bibnamefont{Kusnezov}},
  \bibinfo{journal}{Phys. Lett. A} \textbf{\bibinfo{volume}{265}},
  \bibinfo{pages}{250} (\bibinfo{year}{2000}).

\bibitem[{\citenamefont{Flach and Mutschke}(1994)}]{flach}
\bibinfo{author}{\bibfnamefont{S.}~\bibnamefont{Flach}} \bibnamefont{and}
  \bibinfo{author}{\bibfnamefont{G.}~\bibnamefont{Mutschke}},
  \bibinfo{journal}{Phys. Rev. E} \textbf{\bibinfo{volume}{49}},
  \bibinfo{pages}{5018} (\bibinfo{year}{1994}).

\bibitem[{\citenamefont{Schneider and Stoll}(1978)}]{stoll}
\bibinfo{author}{\bibfnamefont{T.}~\bibnamefont{Schneider}} \bibnamefont{and}
  \bibinfo{author}{\bibfnamefont{E.}~\bibnamefont{Stoll}},
  \bibinfo{journal}{Phys. Rev. B} \textbf{\bibinfo{volume}{18}},
  \bibinfo{pages}{6468} (\bibinfo{year}{1978}).

\bibitem[{\citenamefont{Morse and Feshbach}(1953)}]{morse}
\bibinfo{author}{\bibfnamefont{P.~M.} \bibnamefont{Morse}} \bibnamefont{and}
  \bibinfo{author}{\bibfnamefont{H.}~\bibnamefont{Feshbach}},
  \emph{\bibinfo{title}{Methods of Theoretical Physics}}
  (\bibinfo{publisher}{McGraw-Hill, New York}, \bibinfo{year}{1953}).

\bibitem[{\citenamefont{Weiss}(2002)}]{weiss}
\bibinfo{author}{\bibfnamefont{G.}~\bibnamefont{Weiss}},
  \bibinfo{journal}{Physica D} \textbf{\bibinfo{volume}{311}},
  \bibinfo{pages}{381} (\bibinfo{year}{2002}).

\bibitem[{\citenamefont{Chaikin and Lubensky}(1995)}]{lubensky}
\bibinfo{author}{\bibfnamefont{P.~M.} \bibnamefont{Chaikin}} \bibnamefont{and}
  \bibinfo{author}{\bibfnamefont{T.}~\bibnamefont{Lubensky}},
  \emph{\bibinfo{title}{Principles of Condensed Matter Physics}}
  (\bibinfo{publisher}{Cambridge University Press, Cambridge},
  \bibinfo{year}{1995}).

\bibitem[{\citenamefont{Mclachlan and Atela}(1992)}]{MA92}
\bibinfo{author}{\bibfnamefont{R.~I.} \bibnamefont{Mclachlan}}
  \bibnamefont{and} \bibinfo{author}{\bibfnamefont{P.}~\bibnamefont{Atela}},
  \bibinfo{journal}{Nonlinearity} \textbf{\bibinfo{volume}{5}},
  \bibinfo{pages}{541} (\bibinfo{year}{1992}).

\bibitem[{\citenamefont{Aoki et~al.}(2006)\citenamefont{Aoki, Lukkarinen, and
  Spohn}}]{Spohn06}
\bibinfo{author}{\bibfnamefont{K.}~\bibnamefont{Aoki}},
  \bibinfo{author}{\bibfnamefont{J.}~\bibnamefont{Lukkarinen}},
  \bibnamefont{and} \bibinfo{author}{\bibfnamefont{H.}~\bibnamefont{Spohn}},
  \bibinfo{journal}{J. Stat. Phys.} \textbf{\bibinfo{volume}{124}},
  \bibinfo{pages}{1105} (\bibinfo{year}{2006}).

\bibitem[{\citenamefont{Dauxois et~al.}(1993)\citenamefont{Dauxois, Peyrard,
  and Bishop}}]{dauxois}
\bibinfo{author}{\bibfnamefont{T.}~\bibnamefont{Dauxois}},
  \bibinfo{author}{\bibfnamefont{M.}~\bibnamefont{Peyrard}}, \bibnamefont{and}
  \bibinfo{author}{\bibfnamefont{A.~R.} \bibnamefont{Bishop}},
  \bibinfo{journal}{Phys. Rev. E} \textbf{\bibinfo{volume}{47}},
  \bibinfo{pages}{684} (\bibinfo{year}{1993}).

\bibitem[{\citenamefont{Omelyan et~al.}(2002)\citenamefont{Omelyan, Mryglod,
  and Folk}}]{PEFRL}
\bibinfo{author}{\bibfnamefont{I.}~\bibnamefont{Omelyan}},
  \bibinfo{author}{\bibfnamefont{I.}~\bibnamefont{Mryglod}}, \bibnamefont{and}
  \bibinfo{author}{\bibfnamefont{R.}~\bibnamefont{Folk}},
  \bibinfo{journal}{Computer Physics Communications}
  \textbf{\bibinfo{volume}{146}}, \bibinfo{pages}{188} (\bibinfo{year}{2002}).

\bibitem[{\citenamefont{Volz et~al.}(1996)\citenamefont{Volz, Saulnier,
  Lallemand, Perrin, Depondt, and Mareschal}}]{Volz96}
\bibinfo{author}{\bibfnamefont{S.}~\bibnamefont{Volz}},
  \bibinfo{author}{\bibfnamefont{J.-B.} \bibnamefont{Saulnier}},
  \bibinfo{author}{\bibfnamefont{M.}~\bibnamefont{Lallemand}},
  \bibinfo{author}{\bibfnamefont{B.}~\bibnamefont{Perrin}},
  \bibinfo{author}{\bibfnamefont{P.}~\bibnamefont{Depondt}}, \bibnamefont{and}
  \bibinfo{author}{\bibfnamefont{M.}~\bibnamefont{Mareschal}},
  \bibinfo{journal}{Phys. Rev. B} \textbf{\bibinfo{volume}{54}},
  \bibinfo{pages}{340} (\bibinfo{year}{1996}).

\bibitem[{\citenamefont{Sokolov and Metzler}(2003)}]{Sokolov03}
\bibinfo{author}{\bibfnamefont{I.~M.} \bibnamefont{Sokolov}} \bibnamefont{and}
  \bibinfo{author}{\bibfnamefont{R.}~\bibnamefont{Metzler}},
  \bibinfo{journal}{Phys. Rev. E} \textbf{\bibinfo{volume}{67}},
  \bibinfo{pages}{010101} (\bibinfo{year}{2003}).

\end{thebibliography}

\end{document}